\documentclass[epjST]{svjour}
\usepackage{graphics}
\begin{document}

\title{Computer Simulations of Charged Colloids in Alternating Electric Fields}

\author{Jiajia Zhou \and Friederike Schmid}

\institute{Institut f\"ur Physik, Johannes Gutenberg-Universit\"at Mainz, 
D55099 Mainz, Germany}

\abstract{
We briefly review recent theoretical and simulation studies of charged colloidal dispersions in alternating electric fields (AC fields). 
The response of single colloid to an external field can be characterized by a complex polarizability, which describes the dielectric properties of the colloid and its surrounding electrical double layer. 
We present computer simulation studies of single spherical colloid, using a coarse-grained mesoscale approach that accounts in full for hydrodynamic and electrostatic interactions as well as for thermal fluctuations.  
We investigate systematically a number of controlling parameters, such as the amplitude and frequency of the AC-fields.
The results are in good agreement with recent theoretical predictions. 
} 

\maketitle

\section{Introduction}
\label{sec:introduction}

Colloidal dispersions have numerous applications in various fields such as chemistry, biology, medicine, and engineering \cite{RSS,Hiemenz_colloid3,Kreuter,Grzelczak2010}.
In an aqueous solution, colloidal particles often acquire surface charges spontaneously, due to several mechanisms: dissociation of a surface group, release of ionic impurities, preferential adsorption of ions from the solution, etc. 
Therefore, using electric fields is an obvious and promising way to assess or control the properties of colloidal dispersions. 

Alternating electric fields (AC fields) can be used to manipulate single or multiple colloidal particles. 
This provides an attractive method to control the position of individual colloid or the structure of colloidal suspensions, because the charged colloids respond to the electric fields on relatively short time scales and in an often fully reversible way.
Compared to static electric fields (DC fields), AC fields have the advantage to avoid the constant flow induced by electroosmosis and the accumulation of charged species on electrodes.
Besides the amplitude, the frequency and the phase of the field can also be tuned to probe selectively the dynamics of colloidal suspension in different time scales. 
This opens ample opportunities to design optimal AC pulses for a given task. 

One particularly important example is dielectrophoresis \cite{Pohl,Jones}. 
In dielectrophoresis, the external electric fields polarize the colloids, which are then in turn driven along the direction of the field gradient. 
The time-averaged force acting on a particle in an AC-field $\mathbf{E} \exp(i\omega t)$ is proportional to $\Re \{ \alpha(\omega) \} \nabla |\mathbf{E}|^2$, where $\alpha(\omega)$ is the complex polarizability of the particle.
The differences in the particle size, shape, density, surface properties, etc. result in different induced dipole moments, which in turn leads to distinct movement of the particle under the external field.
The dielectrophoretic effect can be used for trapping colloidal suspensions in ''electric bottles'' \cite{Sullivan2006,Leunissen2008,Leunissen2008a}, and separating colloids  \cite{Green1997,Gascoyne2002} or biologic matters (DNA, viruses, cells, etc.) \cite{Markx1995,Morgan1999,Regtmeier2007}.
It thus has many potential applications in nano- and biotechnology.

Separation of particles in dielectrophoresis utilizes the spatial inhomogeneity of the AC fields. 
In a homogeneous setup, AC fields can also direct the self-assembly of many particles \cite{Grzelczak2010}. 
The colloidal particles acquire dipole moments under the external fields, and the dipole-dipole interactions between particles can be tuned by adjusting the field strength and frequency, which subsequently enable ones to control the self-assembled structures precisely \cite{Yethiraj2003,Ristenpart2003,Lumsdon2004,McMullan2010,McMullan2012,Beltramo2013}.
Combined with many other controlling mechanisms such as confinement \cite{Gong2001,Gong2002}, dispersity in colloid size \cite{Hoffman2008}, particle shape \cite{Hermanson2001,Singh2009,Kang2010,Kang2013}, surface properties \cite{Gangwal2008a,ZhangLu2012} etc., AC fields can be used to create diverse and robust structures in colloidal dispersions. 

The crucial quantity in applications of AC fields is the polarizability $\alpha(\omega)$.
When an external electric field is applied to the suspension, both the colloidal particle and its surrounding electric double layer will be polarized. 
The colloid acquires a dipole moment of the form $\mathbf{p} \exp(i \omega t)$ in the direction of the applied field, and the amplitude of the dipole moment can be written as
 \begin{equation}
  \label{eq:alpha_def}
  \mathbf{p} = \alpha(\omega) \mathbf{E} .
 \end{equation}
The polarizability $\alpha(\omega)$ characterizes the dielectric response of the colloids, and it is a complex function of the frequency and the amplitude of the external field. 
In the linear response region, the dipole moment is proportional to the magnitude of the external field; thus the polarizability does not depend on the field strength for weak external fields.
For strong electric fields, nonlinear effects may play an important role. 

Two important length scales emerge in discussions of dilute colloidal dispersions: the particle radius $R$ and the Debye screening length $l_D$.
The thickness of the electric double layer (EDL) is characterized by the Debye screening length
 \begin{equation}
  \label{eq:Debye}
  l_D = \left[ 4 \pi l_B \sum_i z_i^2 \rho_i(\infty) \right]^{-1/2},
 \end{equation}
where the summation runs over different ion types. 
In Eq.~(\ref{eq:Debye}), $l_B=e^2/(4\pi\epsilon_{\rm m} k_BT)$ is the Bjerrum length which depends on the medium permittivity $\epsilon_{\rm m}$ and the temperature of the solution $T$, and $z_i$ and $\rho_i(\infty)$ are the valence and bulk concentration for type $i$ ion, respectively.
We shall see late these two length scales determine two important time scales in the colloidal response to the external fields.

In this paper, we review some basic concepts of the theoretical description and recent progress in simulation studies of colloidal dispersions under alternating electric fields. 
We focus on electrokinetic theories for the polarizability of a single colloidal particle and on mesoscopic simulation methods which are used to test the theory. 
We will describe recent results from Dissipative Particle Dynamics simulations and compare them to theoretical predictions.

The minireview is organized as follows: we start with a discussion of the analytic theories and numerical solutions based on the standard electrokinetic model in section~\ref{sec:theory}.
Then we turn to mesoscopic simulations and describe various model that can be used for mesoscale simulations of charged colloids in electrolyte solution in section~\ref{sec:model}.
We present our simulation results in section~\ref{sec:dpd}, which are based on the Dissipative Particle Dynamics and a raspberry model of the colloid.
Finally, we conclude in section~\ref{sec:summary} with a brief summary.

\section{Theoretic Background}
\label{sec:theory}

\subsection{Maxwell-Wagner-O'Konski theory}

We consider the simplest case of a single spherical particle of radius $R$ (with permittivity $\epsilon_{\rm p}$ and conductivity $K_{\rm p}$) immersed in an electrolyte solution medium (with permittivity $\epsilon_{\rm m}$ and conductivity $K_{\rm m}$).
We focus on the dielectric response of the colloid, \emph{i.e.}, the polarizability $\alpha(\omega)$, under the presence of an external AC field of the form $E \cos(\omega t) \hat{\mathbf{x}}$. 
The frequency-dependence of the dipole moment stems down to the fact that there is property mismatch at the solid/liquid interface.
This is widely known as the Maxwell-Wagner mechanism of dielectric dispersion \cite{Maxwell1954,Wagner1914}. 
The polarizability can be written in terms of the Clausius-Mossotti factor $K$,
\begin{equation}
  \label{eq:alpha_mw}
  \alpha(\omega) = 4\pi \epsilon_{\rm m} R^3 K(\omega) = 4\pi \epsilon_{\rm m} R^3 \frac{ \epsilon_{\rm p}^* - \epsilon_{\rm m}^* }{ \epsilon_{\rm p}^* + 2 \epsilon_{\rm m}^*},
\end{equation}
where $\epsilon_{\rm p}^*$ and $\epsilon_{\rm m}^*$ are the complex dielectric constants of the particle and the medium: $\epsilon_{\rm p,m}^* = \epsilon_{\rm p,m} + K_{\rm p,m}/i\omega$. 
In the low frequency limit, we obtain $K(\omega \rightarrow 0) = (K_{\rm p} - K_{\rm m})/(K_{\rm p} + 2K_{\rm m})$,
thus the response is dominated by the conducting properties of the system.
In the high frequency limit, we reach a different result $K(\omega \rightarrow \infty) = (\epsilon_{\rm p} - \epsilon_{\rm m} )/( \epsilon_{\rm p} + 2\epsilon_{\rm m} )$, indicating that the dielectric properties determine the high frequency behavior. 
This contrast originates from the fact that the electric field propagate fast and can respond to external perturbation almost instantaneously, whereas charges can not.  

The frequency which separates the low and high frequency regions is the inverse of the Maxwell-Wagner relaxation time
\begin{equation}
  \label{eq:tmw}
  \tau_{\mathrm{mw}} = \frac{\epsilon_{\rm p} + 2\epsilon_{\rm m}}{K_{\rm p} + 2K_{\rm m}}.
\end{equation}
For the special case discussed below ($\epsilon_{\rm m}=\epsilon_{\rm p}$, $K_{\rm p}=0$ and 1-1 electrolytes), we can rewrite Eq.~(\ref{eq:tmw}) using the conductivity formula $K_{\rm m}=2\rho_s e^2 D_I/k_BT$, where $\rho_s$ is the bulk salt concentration and $D_I$ is the diffusion constant of microions.
The result is 
\begin{equation}
  \tau_{\rm mw} = \frac{3}{2} \frac{\epsilon_{\rm m}}{K_{\rm m}} = \frac{3}{2} \frac{l_D^2}{D_I}. 
\end{equation}
Thus the physical meaning of $\tau_{\rm mw}$ is the time required for a microion to diffuse over the distance of Debye screening length. 
The finite time required for the formation of the free charges near the colloid surface is in fact responsible for the Maxwell-Wagner dispersion.

In the classical Maxwell-Wagner theory, the polarizability (\ref{eq:alpha_mw}) only depends on the bulk properties of the solution and the colloid, thus it can be applied to both uncharged and charged colloids.
For charged colloids, the electric double layer plays an important role in determining the dielectric response.
Since the microions that form the EDL are mobile, they can also move under the external fields. 
The contribution from EDL can be included in the Maxwell-Wagner theory as
a surface conductance term, which was first proposed by O'Konski
\cite{OKonski1960}. 
The Maxwell-Wagner-O'Konski theory was extended to ellipsoidal colloids \cite{Saville2000}, and had been used to interpret experimental results \cite{Green1999,Ermolina2005}.

\subsection{Low-frequency dielectric response}

When an external electric field is applied to a charged colloid (assumed to be negatively charged), the positive counterions in the EDL migrate in the direction of the field along the surface of the colloid.
This effectively creates an excess of counterions at the front end of the colloid, and a deficit at the back end. 
As a response to this imbalance, counterions are withdrawn from the EDL to the bulk in one end, while in the opposite end, counterions are pushed from the bulk into the EDL.
On long time scales, a salt concentration gradient builds up along the colloid and the thickness of the double layer varies accordingly, leading to an additional source of polarization. 
The characteristic time scale for building up the concentration gradient is basically given by the time required for a microion to diffuse over the distance of the colloid diameter,
\begin{equation}
  \label{eq:tc}
  \tau_{\rm c} = \frac{(2R)^2}{D_I}.
\end{equation}
In the Maxwell-Wagner-O'Konski theory, the effect of the counterion migration in the EDL is approximated using a surface conductance, but the contribution due to the microion diffusion in and out EDL is ignored.
Theories that take into account the concentration polarization in the low-frequency regime have been developed in the Ukraine school \cite{DukhinShilov,Dukhin1993,Grosse1996}.
The so-called Dukhin-Shilov theory is based on the standard electrokinetic model \cite{RSS} and assumes that the EDL is at local equilibrium with the surrounding bulk solution.
The analytic formalism also relies on the assumption that the EDL is much thinner than the colloid radius. 
This is justified for micrometer-sized colloids, but becomes questionable for particles of nanometer radius. 
The low-frequency dielectric response theory has been extended to asymmetric electrolytes \cite{Hinch1984,Chassagne2003,Grosse2009a,Grosse2009b} and for aspherical colloids \cite{Chassagne2008}.
Alternatively, Dhont and Kang had developed theories for special cases of weakly and strongly charged rod-like colloids \cite{Dhont2010,Dhont2011}. 

In situations that involve thick EDL and the whole frequency spectrum, one can solve the full electrokinetic equations using a variety of numerical methods \cite{DeLacey1981,Fixman1983,Mangelsdorf1992,Lopez-Garcia1996,Mangelsdorf1997,Hill2003,Hill2005,ZhouHao2005,Kim2006,Nakayama2008,Zhaohui2009,Schmitz2012}.
In a seminal paper \cite{DeLacey1981}, DeLacey and White presented numerical solutions to the polarizability of a single colloid under AC fields. 
The method was later extended to include the inertia terms in the Navier-Stokes equations \cite{Mangelsdorf1992,Mangelsdorf1997,ZhouHao2005}.
Hill \emph{et al.} had developed an alternative scheme which overcomes the numerical instability at high-frequencies \cite{Hill2003,Hill2005}.
The numerical methods are also extended to study the polarization of nanorods \cite{ZhaoHui2010} and colloids with slippery surfaces \cite{Khair2009,ZhaoHui2010a}. 
The effect of the colloid concentration was also investigated using a cell model approximation \cite{Carrique2008a,Carrique2008b,Roa2012}.

\section{Simulation Models}
\label{sec:model}

For a system that includes large colloids and small solvent/microions, both the electro-static and hydrodynamic interactions play important roles.
From a computational point of view, the main difficulty in modelling such a system lies in the fact that both the electrostatic and hydrodynamic interactions are long-ranged \cite{Pagonabarraga2010,Rotenberg2013}. 
The long-ranged nature indicates that there are coupling between very different length scales in structure and many time scales in dynamics.
One can use the atomistic Molecular Dynamics (MD) simulations, but accessing large length and time scales comes with enormous requirements regarding computer resources.
In recent years, a number of coarse-grained simulation methods have been developed to address this multi-scale problem. 
The general idea is to couple a MD model of a colloidal particle with a mesoscopic model for Navier-Stokes fluids. 
There are a few choices of the fluid model in the literature, such as the
Lattice Boltzmann (LB) method \cite{Succi,Raabe2004,Yeomans2006,Duenweg2009}, 
Dissipative Particle Dynamics (DPD)
\cite{Hoogerbrugge1992,Espanol1995,Groot1997}, and Multi-Particle Collision Dynamics (MPCD) \cite{Malevanets1999,Kapral2008,Gompper2009}.
A review of different mesoscale simulation methods for electrolytes containing macroions, which also addresses the issue of confinement and boundary conditions, can be found in Ref.~\cite{Smiatek2012}.
In the following, we briefly review several colloid models and the coupling schemes to the mesoscopic Navier-Stokes solvers.

Lattice Boltzmann method solves a linearized Boltzmann equation in a fully discretized fashion. 
In a first attempt, Ladd \emph{et al.} modeled the colloid particle as an extended hollow sphere, and implemented the bounce-back collision rules to realize the no-slip boundary condition \cite{Ladd1993,Ladd1994a,Ladd1994b,Ladd1995,Ladd2001}.
The microions are taken into accounts as scalar fields on the lattice within the Poisson-Boltzmann level \cite{Horbach2001,Capuani2004,Rotenberg2010,Giupponi2011,Giupponi2011a}.
An alternative approach is the raspberry model which presents the colloid as a collection of the surface beads.
The coupling to the fluids is implemented using the force-coupling method \cite{Ahlrichs1998,Ahlrichs1999}.
The integrity of the surface beads is maintained either by springs \cite{Lobaskin2004,Lobaskin2004a,Lobaskin2007,Semenov2013} or by fixing the bead position with respect to the colloid center \cite{Chatterji2005,Chatterji2007,Chatterji2010}.

\begin{figure}[htp]
  \centering
  \resizebox{0.61\columnwidth}{!}{%
    \includegraphics{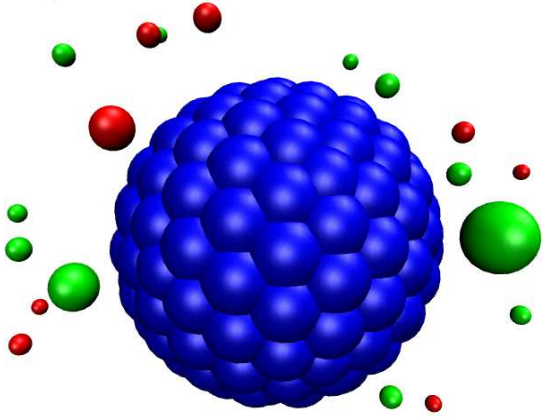} }
  \caption{(Colour on-line) Snapshot of a colloidal particle in a salt solution. The surface sites are represented by the blue beads. The red and green beads are positive and negative microions. Solvent beads are not shown for clarity. Visualization is made by VMD \cite{VMD}.}
  \label{fig:0}
\end{figure}

Dissipative Particle Dynamics is a particle-based simulation method which is Galilean invariant and momentum-conserved. 
The dynamics of the DPD beads consists of alternating streaming and collision steps.
The collision steps are determined from pair interactions between DPD beads, which are coarse-grained and include a dissipative friction term and a random thermal noise. 
In the origin setup \cite{Koelman1993,Boek1997,Steiner2009}, the colloids are made of the same DPD beads as the fluid.
The relative motion of those beads is frozen and the dynamics of the assembly is governed by the movement of the centers of mass and the rotation dynamics of the body axes. 
Espa\~nol proposed the fluid particle model (FPM) \cite{Espanol1998,Dzwinel2000,Pan2008} which treats the colloid as one single object instead of combination of small particles. 
To model the large colloids, two additional non-central shear components are incorporated into the dissipative forces.
We have developed a new DPD model of colloids where the colloid is made of a spherical shell of DPD beads (see Figure \ref{fig:0}) \cite{2012_ac} . 
The fluid beads are prevented from penetrating the colloid by a short-ranged repulsive force. 
One advantage of this model is that the interaction between the surface beads and fluids can be varied to implement different boundary conditions. 
The method generalizes the tunable boundary condition for flat surfaces \cite{Smiatek2008}, which has been successfully to model electrolytes flows near slippery flat surfaces \cite{Smiatek2009,Smiatek2010,Smiatek2011,2012_stripes}.
This approach to modeling slip effects has proved very powerful, it can be used to investigate flows on patterned surfaces with laterally inhomogeneous surface slip \cite{2012_stripes}, and the transport and separation of particles in such laterally inhomogeneous microfluidic channels \cite{Meinhardt2012}.

Similar to DPD, Multi-Particle Collision Dynamics also consists of alternating streaming and collision steps in an ensemble of point particles.
In MPCD, the collision steps are performed by sorting particles in cells and followed by local operations which conserve mass, momentum and energy. 
The coupling between the immersed colloids and the solvent can be implemented by additional forces \cite{Malevanets2000,Lee2001}, bounce-back rules or thermal wall boundary condition \cite{Lamura2001,Lamura2002,Lee2004a,Padding2005,Padding2006,Whitmer2010}.
The presence of solid surfaces introduces slight complication when the bounce-back reflections are used, which can be overcome by using the ``ghost'' or ``wall'' particles \cite{Lamura2001,Lamura2002}.

\section{DPD Simulation}
\label{sec:dpd}

In this section, we briefly review the results from a series of DPD simulations \cite{2012_ac,2012_q0,2013_response} on the response of a spherical particle in salt solution to an applied AC field, focussing on the effect of field strength and field frequency. 
In the following, physical quantities will be reported in a model unit system of $\sigma$ (length), $m$ (mass), $\varepsilon$ (energy), $e$ (charge) and a derived time unit $\tau=\sigma\sqrt{m/\varepsilon}$.
We shall consider a special case of nonconducting colloids ($K_{\rm p}=0$) and assume that the colloid and the solution have the same dielectric constant ($\epsilon_{\rm m}=\epsilon_{\rm p}$).

\subsection{Effect of field strength}

\begin{figure}[htp]
  \centering
  \resizebox{0.75\columnwidth}{!}{%
    \includegraphics{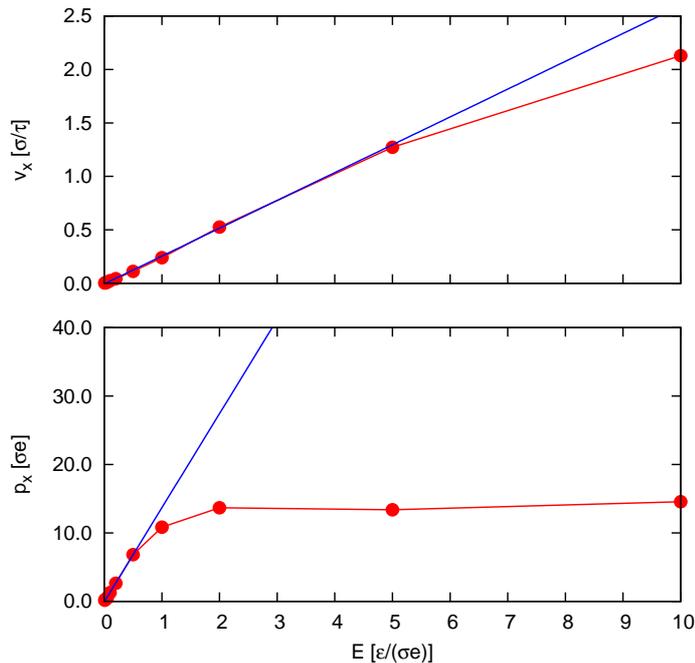} }
  \caption{The amplitude of the colloid velocity (top) and the dipole moment (bottom) as a function of the field strength. The frequency of the applied field is set at $f=0.01\,\tau^{-1}$. The colloid has a radius of $R=3.0\,\sigma$ and total charge $Q=50\,e$. The salt concentration of the solution is $\rho_s=0.05\,\sigma^{-3}$.}
  \label{fig:1}
\end{figure}

We begin by examining the effect of the field strength in order to determine the validity region of the linear response theory.
Theoretic studies on the dielectric response often assume that external fields are weak; thus the governing equations can be linearized to make analytical process. 
In simulations, large electric fields can be studied to examine the effect of the field strength and possible nonlinear effects. 
Figure~\ref{fig:1} shows the amplitudes of the colloid velocity and dipole moment as a function of the field strength. 
The simulations are performed at a low frequency $f=0.01\,\tau^{-1}$.
The field strength is varied over two orders of magnitude from $0.01$ to $10\,\varepsilon/(\sigma e)$.  

For weak fields, the velocity is a linear function of the field strength up to $E \approx 2.0\, \varepsilon / (\sigma e)$, corresponding to a constant mobility. 
Similarly, the linear dependence of the dipole moment on the field strength indicates a constant polarizability, but the deviation from the linear regime sets in around $E \approx 1.0\, \varepsilon / (\sigma e)$. 
 
For higher field strength, the growth of the velocity slows down, which implies a decreasing mobility.  
This is in contrast to the case of DC fields, where the mobility is found to increase above the linear region \cite{Lobaskin2004a}.  
One possible explanation is that AC fields of large field strength can excite colloidal motion at various frequencies besides the frequency of the applied field. 
Looking at the dipole moment, one finds that deviation from the linear dependence is much more pronounced.  
This is due to the fact that the dipole moment indicates how ions are distributed around the colloidal particle, while the microions in the EDL are mobile and can be easily perturbed by the external field.
For low and intermediate field strength, the main effect of the external field is to elongate the ion clouds in the field direction.  
When the field is strong enough, it may rip off microions which are originally bound to the colloidal surface, which results in a reduction in the dipole moment.

\subsection{Effect of field frequency}

Next we discuss the effect of varying the frequency of external AC fields.  
The amplitude is chosen in the linear region, $E=0.5 \,\varepsilon/(\sigma e)$.  
The frequency is varied over three orders of magnitude from $f=10^{-3}$ to $2. 0\, \tau^{-1}$.  

\begin{figure}[htbp]
  \centering
  \resizebox{0.8\columnwidth}{!}{%
  \includegraphics{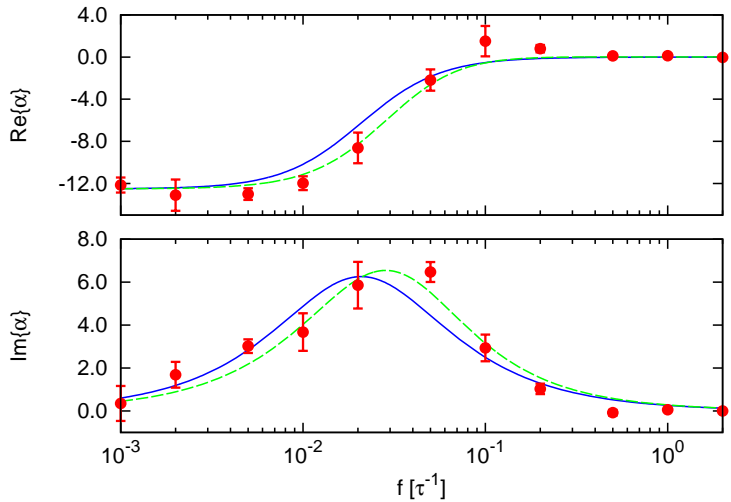} }
  \caption{Real and imaginary part of the complex polarizability $\alpha(\omega)$ of an uncharged particle as a function of the frequency of applied electric field. The field strength is set in the linear region $E=0.5 \, \varepsilon / (\sigma e)$. The salt concentration in the solution is $0.0125 \sigma^{-3}$. The points with err-bars are simulation results. The solid lines give the prediction from the Maxwell-Wagner theory. The dashed line shows the results of Dhont and Kang \cite{Dhont2010}.}
  \label{fig:2}
\end{figure}

We start with the simple case of an uncharged particle, where the absence of the EDL simplifies the picture considerably.  
Figure \ref{fig:2} shows the polarizability for an uncharged particle in a solution with a salt concentration $0.0125 \, \sigma^{-3}$.  
The top figure shows the real part of the polarizability ${\rm Re} \{ \alpha \}$, which is the in-phase component of the dipole moment with respect to the external field.
The bottom figure shows the imaginary part ${\rm Im} \{ \alpha \}$, which is the out-of-phase contribution.  
The real part in the low-frequency limit is negative, thus the induced dipole is pointing in the opposite direction of the external field.  
Since uncharged colloids are not surrounded by an EDL, the negative dipole moment is induced solely by the motion of the salt ions and the obstacle effect of the colloid.   

The solid line in Figure~\ref{fig:2} shows the prediction from the Maxwell-Wagner theory (Eq. \ref{eq:alpha_mw}), which is in good agreement with the simulation. 
The theory predicts a transition between two regimes which is recovered in the simulation at roughly the predicted transition frequency. 
The dashed curves in Figure~\ref{fig:2} show the results of Dhont and Kang \cite{Dhont2010}.
The better agreement to the simulation results is due to the fact that Dhont-Kang theory takes into account the spatial variation of the polarization charges. 
The polarization charges are distributed over a characteristic length of Debye screening length. 
For high salt concentrations, the Debye screen length is much smaller than the colloid size, thus the contribution due to the distribution of polarization charges is small. 
At the other limit of low salt concentration, the contribution becomes more significant, which results in a slight increase of the transition frequency.
Note that for uncharged colloids, there is only one characteristic frequency which is the inverse of the Maxwell-Wagner relaxation time (Eq. \ref{eq:tmw}).

\begin{figure}[htp]
  \centering
  \resizebox{0.8\columnwidth}{!}{%
    \includegraphics{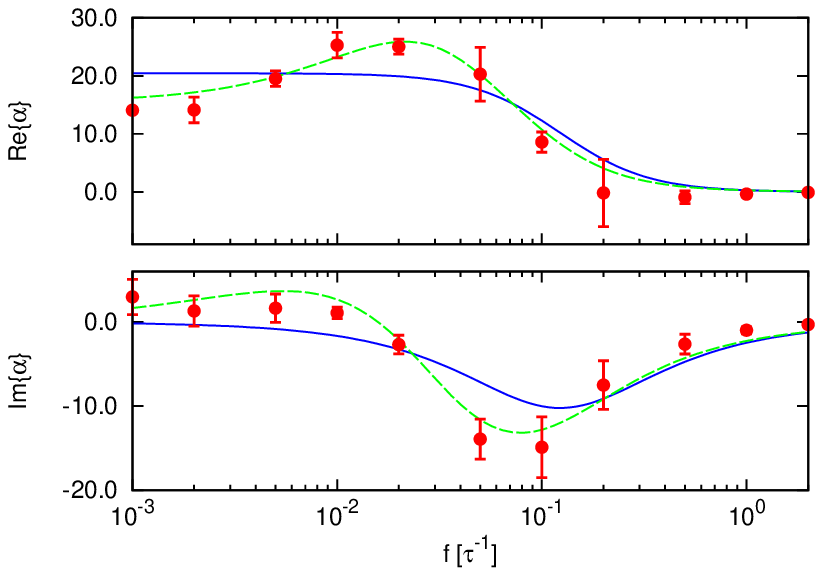} }
  \caption{Real and imaginary part of the complex polarizability $\alpha(\omega)$ of a charged particle ($Q=50\, e$) as a function of the frequency of applied electric field. The field strength is set in the linear region $E=0.5 \, \varepsilon / (\sigma e)$. The salt concentration in the solution is $0.0125 \sigma^{-3}$. The points with err-bars are simulation results. The solid lines give the prediction from the Maxwell-Wagner-O'Konski theory. The dashed lines are numerical solutions to the electrokinetic equations~\cite{Hill2005}.}
   \label{fig:3}
\end{figure}

The situation is more interesting for charged colloids due to the
presence of EDL.
Figure \ref{fig:3} shows the polarizability for a charged colloid ($Q=50\,e$) in a salt solution $\rho_s=0.0125\, \sigma^{-3}$.  
In the low-frequency region, the external perturbation is relatively slow so the colloid and the surrounding EDL can response, resulting in a dipole moment which is in phase with the external field.
The real part ${\rm Re}\{\alpha\}$ has a positive value, in contrast to the negative value for uncharged colloids, so the direction of the induced dipole moment is the same as the applied field.  
Without external fields, the centers of the counterion cloud and the spherical particle coincide, thus the total dipole moment is zero. 
With external fields, the positively charged colloid moves in the direction of the field, while the negatively charged ion cloud moves in the opposite direction.  
The resulting net dipole moment is in the same direction as the external field, therefore one has ${\rm Re}\{\alpha\}>0$.  
In the high-frequency region, the change of the external fields is too fast for the colloid and the ion cloud to response, thus both ${\rm Re}\{\alpha\}$ and ${\rm Im}\{\alpha\}$ reduce to zero. 
At intermediate frequencies, the real part ${\rm Re}\{\alpha\}$ crosses over from positive to zero.  
Below the transition frequency there is an overshoot and after the transition frequency a small undershoot to negative values.  
The imaginary part ${\rm Im}\{\alpha\}$ reaches a maximum at the transition, indicating that the response is out of phase.

The solid lines in Figure~\ref{fig:3} show the prediction from Maxwell-Wagner-O'Konski theory, where the surface conductance due to the EDL is calculated using Bikerman's expression \cite{Bikerman1940,OBrien1986}.
The results are only in qualitative agreement with the simulation.
The theory captures the main features, and roughly the correct transition frequency (Eq.~\ref{eq:tmw}) from the low to high frequency regions, but it misses many details in the intermediate frequencies.  
To take into account the contribution from the EDL, one can solve the full electrokinetic equations using numerical methods. 
The numerical solutions obtained using program MPEK \cite{Hill2005} are shown in dashed curves, which are in good agreement with the simulation.
In particular, the results capture the dielectric dispersion at low frequency (Eq.~\ref{eq:tc}), which is due to the buildup of the salt concentration gradient.

\section{Conclusions}
\label{sec:summary}

In this review, we have examined recent theoretical and simulation studies of dielectric response of dilute colloid dispersions. 
We reviewed general theories for the dipole moment induced by external AC-fields. 
The main difficulty is to include the contribution from EDL properly. 
For EDLs that are thin in comparison to the colloid size, the Maxwell-Wagner-O'Konski theory and the Dukhin-Shilov theory can be used to interpret experimental results.
For EDLs with arbitrary thickness, one can solve the electrokinetic equations using numerical methods.
We also reviewed the several colloid models implemented in mesoscopic simulations, with a focus on the dissipative particle dynamics simulations.
Mesoscale simulation data for spherical particles are in good agreement with the predictions of the electrokinetic theory.
In the future, mesoscale simulations can also be used to study more complex situations, e.g., nonspherical colloids and interacting colloids, where theoretical descriptions are difficult or not available.

We conclude with a brief outlook on potential future directions for the simulation of charged colloids.
Here we have focused on the case of dilute suspensions. 
In a dense suspension or at low salt concentration, the Debye length becomes comparable to the particle spacing. 
The EDLs of neighbouring particles may overlap, and the polarizability will become dependent on the colloid density. 
To explain or predict the self-Assembly process in dense suspensions, a better understand of the effective interaction between particles is required. 
Another possibility is to place the colloids close to a surface or in between surfaces. 
The interaction between the particle and the surface with tunable properties will open new possibilities for manipulating particles, which may be used for particle separation or self-assembly.

\begin{acknowledgement}
We are grateful to Prof. Reghan Hill of McGill University for providing the MPEK program.
We have benefited from discussions and collaborations with Burkard D\"unweg, Vladimir Lobaskin, Stefan Medina Hernando, Sebastian Meinhardt, Taras Molotilin, Roman Schmitz, Jens Smiatek, Olga Vinogradova, and Peter Virnau.
This work was supported by the DFG within SFB TR6 (project B9).
Computational resources at John von Neumann Institute for Computing (NIC J\"ulich), High Performance Computing Center Stuttgart (HLRS Hermit) and JGU Mainz (Mogon Cluster) are gratefully acknowledged.
\end{acknowledgement}

\bibliography{ac}

\end{document}